\documentclass[twocolumn,aps,pra,showpacs]{revtex4}
\usepackage{epsfig}
\begin{document}
\title{Quantum non-demolition measurement saturates fidelity trade-off}
\author{Ladislav Mi\v{s}ta Jr. and Radim Filip}
%\footnote{email:filip@optnw.upol.cz, tel:+420-68-5631572,
%fax:+420-68-5224246}
\affiliation{Department of Optics, Palack\' y University,\\
17. listopadu 50,  772~07 Olomouc, \\ Czech Republic}
\date{\today}
\begin{abstract}
A general quantum measurement on an unknown quantum state enables us 
to estimate what the state originally was. Simultaneously, the
measurement has a destructive effect on a measured quantum state which is 
reflected by the decrease of the output fidelity. 
We show for any $d$-level system that quantum non-demolition 
(QND) measurement controlled by a suitably prepared ancilla is 
a measurement in which the decrease of the output fidelity is minimal. 
The ratio between the estimation fidelity and the output fidelity 
can be continuously controlled by the preparation of the 
ancilla. Different measurement strategies on the ancilla are
also discussed. Finally, we propose a feasible scheme of such a measurement for 
atomic and optical $2$-level systems based on basic controlled-NOT gate.     
\end{abstract}

\maketitle

Measurement in quantum mechanics changes drastically measured quantum
state. Moreover, this change cannot be done arbitrarily small.
This main feature of quantum measurement can be simply proved by 
performing the estimation of the state after the measurement. At first sight this is 
a negative effect which does not allow many operations well known from classical physics. 
Fortunately, there is also a positive aspect of this property. In principle, 
it can be exploited to make communication between two distant stations secure 
against eavesdropping attacks. Namely, secret information can be sent
by quantum states in such a way that any measurement on the transmitted states can 
be detected and consequently any attack on the link can be revealed \cite{Bennett_84}. 
This property represents a fundamental distinction between quantum measurement and classical 
measurement that can be made in principle state non-destructive. Such
an ideal classical measurement has a quantum analogue called {\em quantum non-demolition} 
(QND) measurement \cite{Braginsky_80}. The QND measurement is non-destructive 
in the sense that is preserves probabilistic distribution of so called {\em non-demolition} 
variable of the measured system and simultaneously, the measurement results give a 
perfect copy of the non-demolition variable statistics. From this point of view, 
they can be used as a perfect distributor of information encoded in the non-demolition 
variable of a quantum state. All noise arising in the measurement process is transfered to the
complementary variables. The present work is devoted to (1) the analysis of 
the fundamental property of the QND measurement and (2) to the feasible application 
of the QND measurement for optimal distribution of information encoded in an unknown 
system variable.  

Suppose Alice is given a $d$-level quantum system $S$ (qu$d$it S) in an 
{\it unknown} pure state $|\psi\rangle_{S}$ and she sends the state to Bob. 
Suppose there is Eve between Alice and Bob that wants to guess this state whereas 
disturbing it to the least possible extent. For this purpose Eve can measure the 
state directly by a projective measurement and based on the outcomes of the 
measurement she can guess the state. Alternatively, Eve can guess the state 
from measurement on an ancillary system that previously interacted with the original 
state. Both the strategies produce two states, an estimate of the original state 
that is hold by Eve $\rho_{\rm est}$ and an output state $\rho_{\rm out}$ 
after the measurement that continues toward Bob. The quality of Eve's 
guesses can be characterized by the mean fidelity $G$ (estimation fidelity) 
defined as $G=\int\langle\psi|\rho_{\rm est}|\psi\rangle d\psi$ where
$\int d\psi$ is the integral over the space of pure states and $d\psi$ is the 
measure invariant with respect to unitary transformations.
The perturbation introduced by Eve to the original state can be characterized by the mean 
fidelity $F$ (output fidelity) of the output state $F=\int\langle\psi|\rho_{\rm out}|\psi\rangle d\psi$. 
According to the laws of quantum mechanics the fidelities $F$ and $G$ must satisfy the following 
inequality \cite{Banaszek_01}:
%%%%%%%%%%%%%%%%%%%%%%%%%%%%%%%%%%%%%%%%%%%%%%%%%%%%%%%%%%%%%%%%%%%%%%%%%%%%%%%%%%%%%%%%%%%%%%
\begin{eqnarray}\label{Banaszekd}
\sqrt{F-\frac{1}{d+1}}\leq\sqrt{G-\frac{1}{d+1}}+\sqrt{(d-1)\left(\frac{2}{d+1}-G\right)}.\nonumber\\  
\end{eqnarray}
%%%%%%%%%%%%%%%%%%%%%%%%%%%%%%%%%%%%%%%%%%%%%%%%%%%%%%%%%%%%%%%%%%%%%%%%%%%%%%%%%%%%%%%%%%%%%%
The inequality sets a tightest bound between the mean fidelity $G$ of estimation of an unknown state 
from a general deterministic quantum operation on a single qu$d$it and the mean fidelity 
$F$ of the state after the operation. Particularly important are quantum operations that saturate 
the inequality (\ref{Banaszekd}). Namely, these operations introduce the least possible disturbance 
to the original state in the sense that for a given estimation fidelity $G$ they provide the highest 
possible output fidelity $F$.        

%%%%%%%%%%%%%%%%%%%%%%%%%%%%%%%%%%%%%%%%%%%%%%%%%%%%%%%%%%%%%%%%%%%%%%%%%%%%%%%%%%%%%%%%%%%
\begin{figure}
\centerline{\psfig{width=7.0cm,angle=0,file=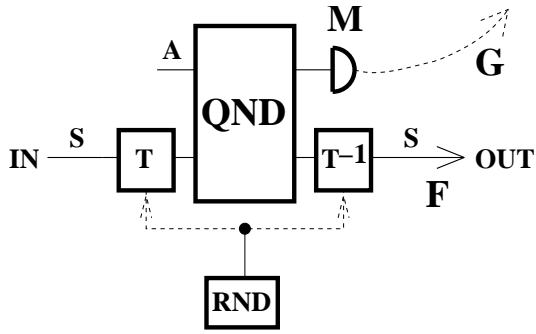}}
\caption{The scheme of optimal measurement with a minimal disturbance: 
QND -- QND interaction, T -- twirling operation, RND -- random-number generator, M -- state discriminator.}
\label{fig1}
\end{figure}
%%%%%%%%%%%%%%%%%%%%%%%%%%%%%%%%%%%%%%%%%%%%%%%%%%%%%%%%%%%%%%%%%%%%%%%%%%%%%%%%%%%%%%%%%%%%

In this article we show generally for a qu$d$it that a perfect QND measurement randomly 
performed along all the basis in the Hilbert space which is controlled by a quantum state 
of ancilla saturates the inequality (\ref{Banaszekd}). In particular, such the QND 
measurement for a single qubit can be implemented by the basic controlled-NOT (CNOT) operation. 
The perfect QND measurement means that Eve has a perfect copy of statistics of the 
non-demolition variable. Further, we discuss in detail an imperfect QND measurement and we compare 
influence of different discriminations of ancillary states. Finally, we shortly summarize 
feasible experimental implementations of the proposed measurement scheme and discuss 
open problems stimulating future research.  

Our protocol depicted in Fig.~\ref{fig1} consists of two steps. At
the outset we consider the protocol without twirling operations $T$
and $T^{-1}$. In the first step the qu$d$it $S$ in an unknown state $|\psi\rangle_{S}$ is coupled by 
the two-qu$d$it unitary interaction $U$ to another ancillary qu$d$it $A$. In the 
second step the information about the state $|\psi\rangle_{S}$ 
is gained from a suitable projective measurement on the ancilla 
and it is converted into the state of another qu$d$it $E$. 
In order the operation to saturate the inequality (\ref{Banaszekd})
the interaction $U$ must satisfy two following conditions.
First, there must exist a normalized vector $|\mu\rangle_{A}$ of the ancilla $A$ such that 
%%%%%%%%%%%%%%%%%%%%%%%%%%%%%%%%%%%%%%%%%%%%%%%%%%%%%%%%%%%%%%%%%%%%%%%%%%%%%%%%%%%%%%%%%%%%%%
\begin{eqnarray}\label{1}
|\psi\rangle_{S}|\mu\rangle_{A}\stackrel{U}{\rightarrow}\sum_{i=1}^{d}c_{i}|a_{i}\rangle_{S}
|\mu_{i}\rangle_{A},    
\end{eqnarray}
%%%%%%%%%%%%%%%%%%%%%%%%%%%%%%%%%%%%%%%%%%%%%%%%%%%%%%%%%%%%%%%%%%%%%%%%%%%%%%%%%%%%%%%%%%%%%%
where $\{|a_{i}\rangle_{S}\}_{i=1}^{d}$ and $\{|\mu_{i}\rangle_{A}\}_{i=1}^{d}$ 
are the orthonormal bases in the state spaces of the qu$d$it $S$ and $A$, respectively, and 
$c_{i}={}_{S}\langle a_{i}|\psi\rangle_{S}$. The interaction (\ref{1})
represents a perfect QND coupling. Second, there must exist a normalized vector 
$|\kappa\rangle_{A}$ that ``switches off'' the interaction $U$, i. e. 
%%%%%%%%%%%%%%%%%%%%%%%%%%%%%%%%%%%%%%%%%%%%%%%%%%%%%%%%%%%%%%%%%%%%%%%%%%%%%%%%%%%%%%%%%%%%%%
\begin{eqnarray}\label{2}
|\psi\rangle_{S}|\kappa\rangle_{A}\stackrel{U}{\rightarrow}|\psi\rangle_{S} 
\frac{1}{\sqrt{d}}\sum_{i=1}^{d}|\mu_{i}\rangle_{A}.    
\end{eqnarray}
%%%%%%%%%%%%%%%%%%%%%%%%%%%%%%%%%%%%%%%%%%%%%%%%%%%%%%%%%%%%%%%%%%%%%%%%%%%%%%%%%%%%%%%%%%%%%%
Obviously, performing the measurement on the ancilla after the
transformation (\ref{2}) in the basis $\{|\mu_{i}\rangle_{A}\}_{i=1}^{d}$ 
gives no information on the input state.   
Now we show that if an unitary interaction $U$ satisfies the two conditions then it can be 
used to construct a quantum operation that saturates the inequality (\ref{Banaszekd}). Moreover, 
the flow of information between the qu$d$it $E$ and the original qu$d$it $S$ can be 
controlled by the preparation of the ancilla $A$. Let us assume that the ancilla is 
prepared in the superposition
%%%%%%%%%%%%%%%%%%%%%%%%%%%%%%%%%%%%%%%%%%%%%%%%%%%%%%%%%%%%%%%%%%%%%%%%%%%%%%%%%%%%%%%%%%%%%%
\begin{eqnarray}\label{ancilla}
|\tau\rangle_{A}=\alpha|\mu\rangle_{A}+\beta|\kappa\rangle_{A},    
\end{eqnarray}
%%%%%%%%%%%%%%%%%%%%%%%%%%%%%%%%%%%%%%%%%%%%%%%%%%%%%%%%%%%%%%%%%%%%%%%%%%%%%%%%%%%%%%%%%%%%%%
where $\alpha$ and $\beta$ are positive real numbers satisfying the normalization condition 
$\alpha^{2}+\beta^{2}+2\alpha\beta/\sqrt{d}=1$. The interaction $U$ transforms the state as
%%%%%%%%%%%%%%%%%%%%%%%%%%%%%%%%%%%%%%%%%%%%%%%%%%%%%%%%%%%%%%%%%%%%%%%%%%%%%%%%%%%%%%%%%%%%%%
\begin{eqnarray}\label{tau}
|\psi\rangle_{S}|\tau\rangle_{A}\stackrel{U}{\rightarrow}
\sum_{i=1}^{d}c_{i}|a_{i}\rangle_{S}\left[\alpha|\mu_{i}\rangle_{A}+
\frac{\beta}{\sqrt{d}}\sum_{j=1}^{d}|\mu_{j}\rangle_{A}\right].   
\end{eqnarray}
%%%%%%%%%%%%%%%%%%%%%%%%%%%%%%%%%%%%%%%%%%%%%%%%%%%%%%%%%%%%%%%%%%%%%%%%%%%%%%%%%%%%%%%%%%%%%%
Then, the ancilla is measured in the basis $\{|\mu_{i}\rangle_{A}\}_{i=1}^{d}$. 
If the ancilla is found in the state $|\mu_{r}\rangle_{A}$ then Eve
prepares her qu$d$it $E$ in the state $|a_{r}\rangle_{E}$. The proposed scheme realizes a general 
quantum operation on qu$d$it $S$ that can be described by 
a set of operators $A_{r}$, ($r=1,\ldots,d$). In the basis 
$\{|a_{i}\rangle_{S}\}_{i=1}^{d}$ the operators are represented 
by diagonal matrices with elements 
%%%%%%%%%%%%%%%%%%%%%%%%%%%%%%%%%%%%%%%%%%%%%%%%%%%%%%%%%%%%%%%%%%%%%%%%%%%%%%%%%%%%%%%%%%%%%%
\begin{eqnarray}\label{A}
(A_{r})_{ij}=\left(\alpha\delta_{ir}+\frac{\beta}{\sqrt{d}}\right)\delta_{ij},   
\end{eqnarray}
%%%%%%%%%%%%%%%%%%%%%%%%%%%%%%%%%%%%%%%%%%%%%%%%%%%%%%%%%%%%%%%%%%%%%%%%%%%%%%%%%%%%%%%%%%%%%%
where $i,j=1,\ldots,d$ and $\delta_{ij}$ is the Kronecker symbol.
Since the operators satisfy the resolution of unity $\sum_{i=1}^{d}A_{r}^{\dag}A_{r}=\openone$
the operation is deterministic and one can use the following formulas \cite{Banaszek_01}:
%%%%%%%%%%%%%%%%%%%%%%%%%%%%%%%%%%%%%%%%%%%%%%%%%%%%%%%%%%%%%%%%%%%%%%%%%%%%%%%%%%%%%%%%%%%%%%
\begin{eqnarray}\label{fidelities}
F&=&\frac{1}{d\left(d+1\right)}\left(d+\sum_{r=1}^{d}|{\rm Tr}A_{r}|^2\right)\label{F},\\
G&=&\frac{1}{d\left(d+1\right)}\left(d+\sum_{r=1}^{d}
{}_E\langle a_{r}|A_{r}^{\dag}A_{r}|a_{r}\rangle_{E}\right).\label{G}
\end{eqnarray}
%%%%%%%%%%%%%%%%%%%%%%%%%%%%%%%%%%%%%%%%%%%%%%%%%%%%%%%%%%%%%%%%%%%%%%%%%%%%%%%%%%%%%%%%%%%%%%
Hence, one obtains using Eq.~(\ref{A}) that 
%%%%%%%%%%%%%%%%%%%%%%%%%%%%%%%%%%%%%%%%%%%%%%%%%%%%%%%%%%%%%%%%%%%%%%%%%%%%%%%%%%%%%%%%%%%%%%
\begin{eqnarray}\label{FG}
F=\frac{1+\left(\alpha+\sqrt{d}\beta\right)^{2}}{d+1},\quad
G=\frac{1+\left(\alpha+\frac{\beta}{\sqrt{d}}\right)^{2}}{d+1}.
\quad \end{eqnarray}
%%%%%%%%%%%%%%%%%%%%%%%%%%%%%%%%%%%%%%%%%%%%%%%%%%%%%%%%%%%%%%%%%%%%%%%%%%%%%%%%%%%%%%%%%%%%%%
Substituting finally these mean fidelities back into the inequality (\ref{Banaszekd}) we find that they 
saturate the inequality. This means that the fidelities (\ref{FG}) lie for any state (\ref{ancilla}) 
of ancilla on the very boundary of the quantum mechanically allowed region defined by the inequality 
(\ref{Banaszekd}). Moreover, by changing continuously the parameter $\alpha$ in this state 
from $1$ to $0$ one can continuously move along the whole boundary from its one extreme 
point $(G_{\rm max},F_{\rm min})=(2/(d+1),2/(d+1))$ to its other extreme point 
$(G_{\rm min},F_{\rm max})=(1/d,1)$. Up to now we have considered the
device in Fig.~\ref{fig1} without twirling operations $T$ and $T^{-1}$. Such a 
scheme is not universal as the output state fidelity $f=\langle\psi|\rho_{\rm out}|\psi\rangle$ 
is dependent on the input state $|\psi\rangle$. In order to obtain a universal device 
where the fidelity $f$ is state independent and therefore
$f=F$ it is sufficient to place the QND interaction in between 
two twirling operations as is depicted in Fig.~\ref{fig1} \cite{Werner_98}.

Interestingly, the controllable optimal quantum operation (\ref{A}) can be implemented 
using the qu$d$it CNOT gate $U_{\rm CNOT}$ defined as $U_{\rm CNOT}|i\rangle_{S}|j\rangle_{A}=|i\rangle_{S}|i\oplus j\rangle_{A}$ where $\oplus$ denotes addition modulo $d$ and $\{|i\rangle_{S,A}\}_{i=1}^{d}$ 
are chosen sets of basis 
states of qu$d$its $S$ and $A$ \cite{Alber_01}. For the CNOT gate the relevant states of the ancilla 
satisfying the conditions (\ref{1}) and (\ref{2}) are $|\mu\rangle_{A}=|0\rangle_{A}$ and 
$|\kappa\rangle_{A}=(1/\sqrt{d})\sum_{i=1}^{d}|i\rangle_{A}$, respectively. Eve then measures the ancilla 
in the basis $\{|i\rangle_{A}\}_{i=1}^{d}$ and prepares the state $|r\rangle_{E}$ if she finds the ancilla 
in the state $|r\rangle_{A}$. Notice, that optimal quantum operation saturating the inequality 
(\ref{Banaszekd}) can be alternatively realized via teleportation scheme with two 
entangled ancillas \cite{Banaszek_01}. 

\begin{figure}
\centerline{\psfig{width=6.0cm,angle=270,file=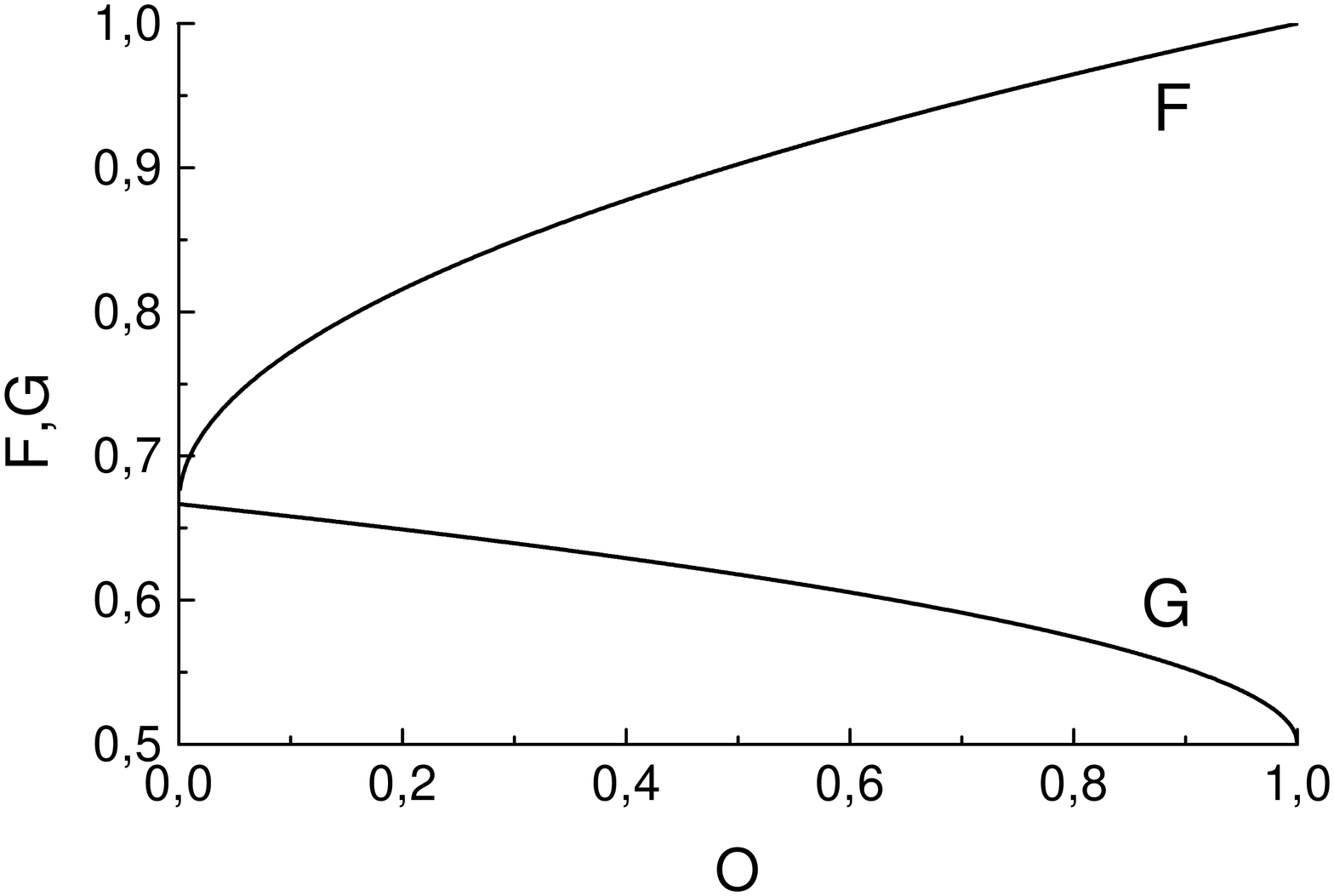}}
\caption{The dependence of the fidelities $F$ and $G$ on the overlap 
$O=|\langle\mu_{1}|\mu_{2}\rangle|^{2}$.}
\label{fig2}
\end{figure}

A specific feature of the quantum operation (\ref{A}) is that it is diagonal in the basis 
$\{|a_{i}\rangle_{S}\}_{i=1}^{d}$ and thus it preserves these basis states. Therefore, 
our scheme can be interpreted as the QND measurement on the qu$d$it $S$ of some observable
(non-demolition variable) ${\cal A}=\sum_{i=1}^{d}a_{i}|a_{i}\rangle_{S}{}_{S}\langle a_{i}|$ 
with non-degenerate eigenvalues $a_{i}$.
In fact, the QND measurement preserving the basis states $\{|a_{i}\rangle_{S}\}_{i=1}^{d}$
can be implemented with a more general class of two-qu$d$it unitary interactions.
To illustrate this, suppose Eve is given an unitary interaction $U$ satisfying only the condition 
(\ref{1}) where, in addition, the states of ancilla $\{|\mu_{i}\rangle_{A}\}_{i=1}^{d}$ are in general 
nonorthogonal. Clearly, in this case Eve's best strategy is to discriminate 
among these states and since she has no a priori information about the occurrence of the states 
(the complex amplitudes $c_{i}$ are unknown) the states have equal a priori probabilities. 
In order to preserve the deterministic character of her operation when in each run 
of the protocol the measurement on the ancilla uniquely determines which of the basis 
states $\{|a_{i}\rangle_{E}\}_{i=1}^{d}$ is to be prepared she has to use an ambiguous 
discrimination \cite{Helstrom_76} of the states $\{|\mu_{i}\rangle_{A}\}$. 
In this approach she applies a generalized measurement $\Pi_{i}$, $i=1,\dots,d$ 
($\Pi_{i}\geq0$,$\sum_{i=1}^{d}\Pi_{i}=\openone$) on the ancilla $A$ discriminating 
among these states and prepares the state $|a_{r}\rangle_{E}$ 
if she detected $\Pi_{r}$. Making use of the Eqs.~(\ref{F}) and (\ref{G}) one then finds that 
%%%%%%%%%%%%%%%%%%%%%%%%%%%%%%%%%%%%%%%%%%%%%%%%%%%%%%%%%%%%%%%%%%%%%%%%%%%%%%%%%%%%%%%%%
\begin{eqnarray}\label{FGn}
F=\frac{2+\frac{1}{d}\sum_{i\ne j=1}^{d}{}_{A}\langle\mu_{i}|\mu_{j}\rangle_{A}}{d+1},\quad
G=\frac{2-P_{e}}{d+1},
\end{eqnarray}
%%%%%%%%%%%%%%%%%%%%%%%%%%%%%%%%%%%%%%%%%%%%%%%%%%%%%%%%%%%%%%%%%%%%%%%%%%%%%%%%%%%%%%%%%%
where $P_{e}=1-(1/d)\sum_{r=1}^{d}{_A}\langle\mu_{r}|\Pi_{r}|\mu_{r}\rangle_{A}$ 
is Eve's error rate. The present quantum operation apparently preserves the basis states 
$\{|a_{i}\rangle_{S}\}_{i=1}^{d}$ and therefore it can be again interpreted as a QND 
measurement. The question that can be risen in this context is whether also this 
QND measurement allows to gain maximum possible information on the input 
state similarly as in the previously discussed scheme. The Eq.~(\ref{FGn}) reveals 
that this is not the case as soon as the states $\{|\mu_{i}\rangle_{A}\}$ are nonorthogonal. 
Namely, Eve's error rate is always greater than zero for nonorthogonal states, i. e. $P_{e}>0$, 
and therefore Eve's estimation fidelity will be always less that the highest possible value
$G_{\rm max}=2/(d+1)$ \cite{Bruss_99}. Another interesting question is whether and 
when the fidelities (\ref{FGn}) saturate the inequality (\ref{Banaszekd}). Since the 
fidelity $F$ in Eq.~(\ref{FGn}) is independent of the measurement on the ancilla 
the corresponding estimation fidelity $G$ will be maximized if Eve performs optimal 
discrimination of the ancillary states $\{|\mu_{i}\rangle_{A}\}_{i=1}^{d}$ that minimizes the error 
rate $P_{e}$. For $d>2$ the minimal error rate can be found only numerically \cite{Jezek_02}. 
However, for qubits ($d=2$) the minimal error rate can be calculated analytically and it is given 
by the formula $P_{e,{\rm min}}=\left(1-\sqrt{1-|\langle\mu_{1}|\mu_{2}\rangle|^{2}}\right)/2$ 
\cite{Helstrom_76}.(In this formula and in what follows we drop the subscript $A$ in states
$\{|\mu_{i}\rangle_{A}\}_{i=1}^{d}$). Hence, using the formula and Eq.~(\ref{FGn}) one obtains the output 
fidelity and the estimation fidelity in the form  
%%%%%%%%%%%%%%%%%%%%%%%%%%%%%%%%%%%%%%%%%%%%%%%%%%%%%%%%%%%%%%%%%%%%%%%%%%%%%%%%%%%%%%%%%%%%%%
\begin{eqnarray}
F&=&\frac{2+|\langle\mu_{1}|\mu_{2}\rangle|\cos\phi}{3},\label{f}\\ 
G&=&\frac{3+\sqrt{1-|\langle\mu_{1}|\mu_{2}\rangle|^{2}}}{6},\label{g}
\end{eqnarray}
%%%%%%%%%%%%%%%%%%%%%%%%%%%%%%%%%%%%%%%%%%%%%%%%%%%%%%%%%%%%%%%%%%%%%%%%%%%%%%%%%%%%%%%%%%%%%%
where $\phi=\mbox{arg}\langle\mu_{1}|\mu_{2}\rangle$. Substituting now the fidelities into 
the inequality (\ref{Banaszekd}) for $d=2$ one finds that they are saturated only if 
$\phi=2k\pi$, where $k$ is an integer. The undesirable phase $\phi$ can be removed 
and thus this optimality condition can be established by applying the phase shift 
$|a_{1}\rangle_{S}\rightarrow|a_{1}\rangle_{S}$ and $|a_{2}\rangle_{S}\rightarrow
e^{-i\phi}|a_{2}\rangle_{S}$ on Bob's qubit after the interaction $U$. 
As a consequence, any two-qubit unitary interaction $U$ satisfying the condition (\ref{1}) 
supplemented by the suitable additional phase shift on a qubit $S$ can be used for the 
QND measurement that saturates the inequality (\ref{Banaszekd}) for $d=2$. 
If, however, the states $\{|\mu\rangle_{i}\}_{i=1}^{2}$ are 
nonorthogonal, it is not possible to achieve the maximum possible 
estimation fidelity $G_{\rm max}=2/3$ and therefore 
one cannot move along the whole boundary of the region defined by 
the inequality (\ref{Banaszekd}) for $d=2$ by changing the state 
$|\mu\rangle_{A}$ of the ancilla.The dependences of the fidelities 
$F$ and $G$ given in Eqs.~(\ref{f}) and (\ref{g}) where $\phi=0$ on
the overlap $O=|\langle\mu_{1}|\mu_{2}\rangle|^{2}$ is depicted 
in Fig.~\ref{fig2}.      
%%%%%%%%%%%%%%%%%%%%%%%%%%%%%%%%%%%%%%%%%%%%%%%%%%%%%%%%%%%%%%%%%%%%%%%%%%%%%%%%%%%%%%%%%%%
%%%%%%%%%%%%%%%%%%%%%%%%%%%%%%%%%%%%%%%%%%%%%%%%%%%%%%%%%%%%%%%%%%%%%%%%%%%%%%%%%%%%%%%%%%%%

However, the maximum possible estimation fidelity can be achieved, at least on a suitable sub-ensemble, 
if Eve replaces the ambiguous discrimination by the unambiguous discrimination \cite{Ivanovich_87} 
of the states $\{|\mu_{i}\rangle\}_{i=1}^{2}$. This strategy can be used 
if the states are linearly independent and it allows to discriminate them perfectly 
with a certain probability $P_{I}$ of inconclusive result. The corresponding 
generalized measurement has three components 
$\Sigma_{1}=|\mu_{2}^{\perp}\rangle\langle\mu_{2}^{\perp}|
/(1+|\langle\mu_{1}|\mu_{2}\rangle|)$, $\Sigma_{2}=|\mu_{1}^{\perp}\rangle\langle\mu_{1}^{\perp}|
/(1+|\langle\mu_{1}|\mu_{2}\rangle|)$ ($|\mu_{i}^{\perp}\rangle$ is 
orthogonal state to the state $|\mu_{i}\rangle$) and 
$\Sigma_{0}=\openone-\sum_{i=1}^{2}\Sigma_{i}$ ($\Sigma_{i}\geq0$). 
The component $\Sigma_{0}$ corresponds 
to the inconclusive result and this measurement is optimal in the sense that the 
probability $P_{I}$ attains minimum possible value 
$P_{I,{\rm min}}=|\langle\mu_{1}|\mu_{2}\rangle|$. Apparently, if Eve detects 
the conclusive result $\Sigma_{i}$, $i=1,2$ then she 
prepares the state $|a_{i}\rangle_{E}$. Therefore, on the sub-ensemble corresponding 
to the conclusive ($C$) results Eve prepares the state 
$\rho_{E,C}=\sum_{i=1}^{2}|c_{i}|^{2}|a_{i}\rangle_{E}{}_{E}\langle a_{i}|$ 
for which the mean estimation fidelity achieves maximum possible value, i. e. 
$G_{C}=G_{\rm max}=2/3$. On the same sub-ensemble, Bob receives
the same mixed state as prepares Eve whence $F_C=2/3$.
The obtained result clearly illustrates that if Eve uses unambiguous
discrimination of the states $\{|\mu_{i}\rangle\}_{i=1}^{2}$ than 
in cases when she detects the conclusive result she is able to obtain 
the best possible estimate of the state $|\psi\rangle_{S}$ even if the
QND interaction is imperfect and encodes the information on the 
state into the nonorthogonal states of ancilla.  

To experimentally test this peculiar property of the QND interaction, the 
experimentalists can use a recent progress in the implementations of the CNOT 
gates between the atoms and photons. Due to a possible long time for manipulations 
of qubits (represented by long-lived electronic states) and high efficiency of 
state detection, trapped and cooled ions are ideally suited for 
implementations of quantum operations. A single-ion CNOT gate has been 
realized some time ago in \cite{Monroe95}. Recently, a two-ion CNOT gate 
based on $^{40}Ca^{+}$ ions in a linear Paul trap which were individually 
addressed using focused laser beams has been implemented \cite{Schmidt-Kaler_03}. 
Also two-ion $\pi$-phase gate demonstrated with $^9Be^{+}$ ions in a harmonic trap 
\cite{Liebfried_03} can be used for the same purpose. Further, probabilistic CNOT gates, 
where the qubits are destroyed upon failure, have been experimentally tested in optical 
systems. Despite of the fact that these CNOT gates for polarization qubits \cite{polQND} 
and path qubits \cite{pathQND} are not deterministic they are sufficient to 
experimentally prove the fundamental trade-off between the estimation fidelity $G$ and 
the output fidelity $F$. To achieve universal character of disturbance introduced 
into the measured state mutually inverse twirling operations \cite{Bennett_96} 
have to be implemented on qubit $S$ before and after the QND interaction 
(see Fig.~\ref{fig1}). In summary, in all the mentioned experimental implementations of 
the CNOT gates the fidelity trade-off can be thus directly experimentally measured.    

There are few important and interesting consequences that have to be noticed. 
They can stimulate broad discussion and future work. First, our result shows 
that it is allowed to achieve any optimal fidelity measurement with a minimal 
disturbance by "programming" the QND interaction by a single program ancillary 
state. This is an interesting result in the context of a previous proof that it 
is not possible to programme any single-qubit unitary operation and measurement 
using only a single qubit program \cite{Nielsen_97,Dusek_02}. More generally, any 
optimal fidelity measurement of a qu$d$it can be programmed by 
an ancilla with the $d$-dimensional Hilbert space. A network for this 
optimal fidelity measurement can be proposed, for example, for qu$d$it with $d=4$ 
as it was suggested in our previous work \cite{Mista_04}. Second, 
the effect of ambiguous discrimination of ancillary states 
outgoing QND interaction has been discussed. As we know there is still 
no bound on maximal success rate for this kind of measurement on a quantum system. 
It is an open question if such the optimal measurement can be also based on 
the QND interaction. Third, in fact we decomposed optimal fidelity measurement with 
minimal disturbance for a single copy into two steps: 
programmable QND coupling and discrimination of ancillary states. 
Thus QND interaction is not only optimal for accessing information encoded 
in single preferred basis but also it is optimal for universal measurement 
without a preferred basis. The difference is only in the twirling operation 
which effectively changes the preferred basis. For many identical copies of input state it is 
an open question if optimal fidelity measurement is based on the same method. 
Can be this strategy used to approach optimal fidelity measurement on many identical 
copies \cite{Banaszek_01b} ? At the end, the problem of the quantum
complementarity and erasure for the QND coupling is closely related
\cite{Englert_96,Ralph_04}. It is known that perfect two-qubit 
QND coupling with arbitrary pure ancillary state is perfectly 
reversible if Eve implements an appropriate measurement and 
Bob performs according to the measurement results an appropriate unitary operation. 
The remaining problem to be discussed is also how the imperfect QND 
coupling can be reversed using only local operations and classical communication. 

%conclusion
In this article, a fundamental property of the QND measurement came to light: 
performing QND measurement randomly along all basis in Hilbert space of the system 
the tightest bound (\ref{Banaszekd}) between the estimation fidelity $G$ and 
the output fidelity $F$ of the measured system can be saturated. 
To change optimal ratio between these fidelities it is sufficient to control 
the ancilla of the QND measurement. Even when the used QND measurement is not 
perfect we can still optimally control the fidelity trade-off but only in a 
restricted range if the output ancillary states are optimally ambiguously 
discriminated. These results are not only important from the fundamental 
point of view but they can be used to distribute information carried by quantum 
state without any preferred basis. 
%%%%%%%%%%%%%%%%%%%%%%%%%%%%%%%%%%%%%%%%%%%%%%%%%%%%%%%%%%%%%%%%%%%%%%%%%%%%%%%%%%%%%%%%%%%%
\medskip
\section*{Acknowledgments}
 
We would like to thank J. Fiur\'{a}\v{s}ek and M. Je\v{z}ek for stimulating 
discussions. The research has been supported by Research Project 
"Measurement and information in Optics" of the Czech Ministry of Education. 
We also acknowledge partial support by the Czech Ministry of Education under 
the project OCP11.003 in the framework of the EU grant COST P11. 
R. F. thanks support by Project 202/03/D239 of the Grant Agency of 
the Czech Republic.
%%%%%%%%%%%%%%%%%%%%%%%%%%%%%%%%%%%%%%%%%%%%%%%%%%%%%%%%%%%%%%%%%%%

\end{document}